\documentclass[12pt]{article}

\usepackage[a4paper,margin=1in]{geometry}
\usepackage{graphicx}
\usepackage{multirow}
\usepackage{booktabs}
\usepackage{url}
\usepackage{hyperref}
\usepackage{amsmath}
\usepackage{setspace}
\usepackage{lineno}
\usepackage{natbib}
\usepackage{caption}
\usepackage{tabularx}
\usepackage{array}
\usepackage{float}
\usepackage[section]{placeins}
\usepackage{authblk}

\hypersetup{
    colorlinks=true,
    linkcolor=blue,
    citecolor=blue,
    urlcolor=blue
}

\title{MUIAnno: An Expert-Annotated Dataset and Evaluation Benchmark for Mobile UI Understanding}

\author[1]{Athar Parvez\thanks{Corresponding author. Email: \texttt{g202393830@kfupm.edu.sa}}}
\author[1]{Muhammad Jawad Mufti}
\author[2]{Muqaddas Gull}
\author[1]{Omar Hammad}

\affil[1]{Information and Computer Science Department, King Fahd University of Petroleum and Minerals, Dhahran 31261, Saudi Arabia}

\affil[2]{SDAIA--KFUPM Joint Research Center for Artificial Intelligence, King Fahd University of Petroleum and Minerals, Dhahran 31261, Saudi Arabia}

\date{}

\begin{document}

\maketitle
\begin{abstract}
Understanding mobile user interfaces is important for building intelligent systems such as automation tools, accessibility solutions, and UI-aware agents. However, progress in this area is still limited by the lack of high-quality datasets that reflect real-world mobile applications and include reliable annotations.
In this work, we introduce \textbf{MUIAnno}, a publicly available expert-annotated dataset\footnote{\url{https://huggingface.co/datasets/atharparvezce/iOS-1K-Mobile-UI-Dataset}} for mobile UI understanding, collected from a diverse set of applications across multiple categories available on the iTunes platform. Each app was manually explored to capture representative UI screens, resulting in a collection that reflects a wide range of layouts and design patterns found in practice. To ensure annotation quality, we developed a custom web-based tool that allows UI/UX experts to label interface elements through a simple drag-and-drop process and generate structured annotations in JSON format.
MUIAnno includes detailed annotations of common UI components such as buttons, input fields, navigation elements, and other key interface elements. In addition to presenting the dataset, we also provide benchmark experiments for UI element detection along with baseline results, offering a starting point for future research.
We believe MUIAnno can support further work in mobile UI understanding and help improve systems that rely on accurate interpretation of interface elements.
\end{abstract}


\section{Introduction}

Mobile applications play a central role in everyday digital interactions, supporting activities such as communication, education, finance, and entertainment. As these applications evolve, their user interfaces (UIs) have become increasingly complex, combining structured layouts with diverse interactive elements. This growing complexity has made automatic UI understanding an important research problem across human-computer interaction, computer vision, and multimodal learning~\citep{deka_rico_2017, chen_ui_2018}. Prior work has shown that UI understanding enables tasks such as design analysis, interface retrieval, and code generation~\citep{chen_wireframe-based_2020, li_screen2vec_2021}, as well as language-based screen description and summarization~\citep{wang_screen2words_2021, leiva_describing_2022}.

Beyond offline analysis, UI understanding is now essential for practical systems that interact directly with software. Applications such as automated GUI testing, accessibility support, and conversational agents rely on accurate interpretation of visual interfaces~\citep{chen_unblind_2020, li_widget_2020}. Recent advances in large language models and vision-language models further emphasize this need, as modern systems are expected to operate on screenshots and respond to natural language instructions~\citep{wang_enabling_2023, li_spotlight_2023}. Models such as ScreenAI demonstrate the potential of multimodal learning for UI understanding, but also highlight the dependence on high-quality annotated data~\citep{baechler_screenai_2024, lee_pix2struct_2023}. However, mobile UI understanding presents unique challenges. Unlike natural images, UIs are structured and semantically dense, consisting of elements such as buttons, icons, text, and navigation components that carry both visual and functional meaning. These elements often appear in complex layouts, making precise localization and interpretation difficult~\citep{li_spotlight_2023, baechler_screenai_2024}. This challenge is particularly critical in applications such as accessibility and testing, where small errors can significantly impact usability~\citep{haque_inferring_2024, yu_vision-based_2025}. Existing datasets have made important contributions but remain limited. The Rico dataset provides large-scale UI data but relies on automatically extracted view hierarchies that may not align with visual content~\citep{deka_rico_2017}. Subsequent work has explored tasks such as widget captioning and screen summarization~\citep{li_widget_2020, wang_screen2words_2021}, while more recent datasets such as MUD and MobileViews focus on scaling data collection through automated pipelines~\citep{feng_mud_2024, gao_mobileviews_2024}. Although these approaches improve coverage, they may introduce noise and inconsistencies, particularly for fine-grained element annotations. This limitation becomes more evident with the rise of multimodal agents and screenshot-based benchmarks. Recent systems are expected to interact with interfaces across mobile and desktop environments~\citep{xie_osworld_2024, li_screenspot-pro_2025}, yet evaluations consistently reveal challenges in grounding and element-level understanding~\citep{wu_os-atlas_2024, qin_ui-tars_2025}. These findings suggest that improving dataset quality is essential for advancing UI understanding systems.

To address these gaps, this study proposes a new expert-annotated dataset for mobile UI understanding, constructed from real-world iOS applications collected through the official iTunes platform. The dataset includes manually explored screens that reflect realistic user interactions and interface diversity. All UI elements are annotated using a custom web-based tool that supports precise bounding box labeling and structured JSON generation, ensuring alignment between visual content and semantic representation. A key focus of this work is annotation quality. The dataset is developed through an expert-driven workflow involving UI/UX professionals and a dedicated verification stage. Approximately 14\% of initial annotations required correction, highlighting the complexity of the task and the importance of systematic validation. This process results in reliable annotations suitable for fine-grained UI understanding.

In addition to dataset construction, this study establishes a benchmark for UI element extraction using multimodal large language models. Models are evaluated in a prompt-based setting, where they interpret UI screenshots and generate structured outputs. This setup reflects realistic usage scenarios and aligns with recent trends toward screenshot-based and agent-oriented UI understanding~\citep{jiang_iluvui_2023, qin_ui-tars_2025}.

To address these gaps, this study makes the following contributions:

(i) We introduce \textbf{MUIAnno}, a new expert-annotated mobile UI dataset consisting of 1,000 screens collected from 38 real-world iOS applications across diverse application categories.

(ii) We develop a custom web-based annotation tool that supports structured, consistent, and fine-grained labeling of UI elements using bounding boxes and semantic labels.

(iii) We provide high-quality UI element annotations validated through a multi-stage expert review process, ensuring semantic consistency and reliable visual grounding.

(iv) We establish a prompt-based benchmark for UI element extraction using multimodal large language models under a unified evaluation setting.

(v) We present an empirical evaluation of closed-source and open-source multimodal models, highlighting their strengths and limitations in fine-grained mobile UI understanding.

The rest of the paper is organized as follows: Section 2 reviews related work on UI datasets and multimodal UI understanding. Section 3 describes the proposed dataset, annotation process, annotation tool, task formulation, and evaluation protocol. Section 4 presents the benchmark experiments, including the task definition, model selection, implementation details, and reproducibility considerations. Section 5 reports and discusses the experimental results. Finally, Section 6 concludes the paper and outlines future research directions.

\section{Related Work}

Research on mobile user interface (UI) understanding has expanded significantly with the rise of intelligent systems that interact with applications through visual and language-based inputs. Prior work in this domain can be broadly categorized into two directions: (1) datasets for UI understanding and (2) multimodal approaches for interpreting UI content.

\subsection{Datasets for UI Understanding}

A wide range of datasets have been proposed to support UI-related tasks. One of the most influential resources is the Rico dataset~\citep{deka_rico_2017}, which provides a large-scale collection of Android UI screens together with view hierarchies extracted from real applications. Rico has enabled substantial progress in data-driven UI modeling and analysis. However, its annotations are derived automatically from accessibility metadata, which may not always align with the actual visual layout of the interface. This limitation reduces its suitability for tasks requiring precise element localization and semantic consistency.

Subsequent work has explored alternative data collection strategies and task formulations. Datasets such as those introduced for widget captioning and screen summarization focus on bridging visual interfaces and natural language descriptions~\citep{li_widget_2020, wang_screen2words_2021}. Similarly, Screen2Vec provides semantic representations of UI screens and components, enabling modeling of screen similarity and interaction patterns~\citep{li_screen2vec_2021}. Other efforts have investigated UI design understanding and retrieval using visual features and representation learning~\citep{chen_wireframe-based_2020}. These datasets and methods broaden the scope of UI understanding but are typically tailored to specific tasks rather than general-purpose element-level annotation. More recent datasets have focused on improving scale and coverage through automated pipelines. MobileViews~\citep{gao_mobileviews_2024} introduces a large-scale collection of mobile screens obtained through automated app traversal, significantly expanding dataset size and diversity. Similarly, MUD~\citep{feng_mud_2024} aims to construct a cleaner and noise-filtered dataset for modern UI modeling as shown in Table~\ref{tab:dataset_comparison}. While these approaches are effective in scaling data collection, they still rely on automated or semi-automated processes, which can introduce noise and inconsistencies in fine-grained annotations. Beyond static UI datasets, recent studies have also examined task-oriented and sequential interaction data. For example, GUIOdyssey focuses on cross-application navigation with reasoning annotations [30], while MONDAY uses large-scale mobile video data to support mobile agent training [21]. Although these datasets are useful for studying interaction and decision-making in mobile environments, they are not intended for precise pixel-level annotation of individual UI elements.

\subsection{Multimodal UI Understanding}

In parallel with dataset development, recent research has increasingly focused on multimodal approaches for UI understanding. Vision-language models and large language models (LLMs) have shown strong capabilities in interpreting UI screenshots and supporting natural language interaction with interfaces~\citep{wang_enabling_2023,li_spotlight_2023}. These models can support tasks such as screen description, UI element identification, and interaction planning without requiring task-specific training. Several studies have extended UI understanding into broader multimodal reasoning frameworks. ScreenAI introduces a vision-language model trained for UI and infographic understanding, showing the value of screen-specific pretraining for this domain~\citep{baechler_screenai_2024}. Similarly, Pix2Struct uses screenshot parsing as a pretraining objective for visual-language understanding across different domains, including user interfaces~\citep{lee_pix2struct_2023}. ILuvUI further explores instruction-tuned multimodal modeling for UI reasoning and conversational interaction~\citep{jiang_iluvui_2023}. More recent work has shifted toward general-purpose GUI agents that can interact with real software environments. OSWorld highlights the difficulty of grounding and executing tasks in real computer interfaces~\citep{xie_osworld_2024}, while OS-ATLAS and UI-TARS study large-scale training and evaluation for GUI agents~\citep{wu_os-atlas_2024,qin_ui-tars_2025}. Recent 2026 studies also continue this direction, with Trifuse focusing on multimodal fusion for GUI grounding~\citep{ma_trifuse_2026} and vision-language diffusion models being explored for GUI grounding and action prediction across web, desktop, and mobile interfaces~\citep{kumbhar_towards_2026}. Together, these studies show that UI perception, spatial grounding, and element-level understanding remain important challenges for agentic UI systems.

Despite this progress, several limitations remain. Many existing datasets rely heavily on automated annotations or focus on specific tasks such as navigation, accessibility, design feedback, or UI critique, rather than providing comprehensive element-level annotations. As a result, they often do not provide the level of precision and semantic consistency needed for fine-grained UI understanding. In addition, the performance of multimodal models is closely tied to the quality of the data used to train and evaluate them. Recent studies show that inaccurate UI annotations and inconsistent dataset construction can affect model reliability, especially for tasks that require precise grounding and structured outputs~\citep{hui_winclick_2025,xie_scaling_2025}. This highlights the need for high-quality, expert-validated datasets that can support reliable evaluation and further progress in mobile UI understanding.

\begin{table}[h]
\centering
\caption{Comparison of existing UI datasets with the proposed dataset}
\label{tab:dataset_comparison}
\begin{tabular}{lcccc}
\toprule
Dataset & Source & Annotation & Expert & Focus \\
\midrule
Rico \citep{deka_rico_2017} & Play Store & Automatic & No & Structure \\
MobileViews \citep{gao_mobileviews_2024} & Screenshots & Semi/Auto & No & UI elements \\
MUD \citep{feng_mud_2024} & Play Store & Semi/Auto & No & Modeling \\
MONDAY \citep{jang_scalable_2025} & Videos & Automatic & No & Tasks \\
GUIOdyssey \citep{lu_guiodyssey_2025} & Multi-app & Structured & No & Navigation \\
UICrit \citep{duan_uicrit_2024} & Screenshots & Human & Yes & Feedback \\
\textbf{MUIAnno} & App Store & Human & Yes & Extraction \\
\bottomrule
\end{tabular}
\end{table}

Overall, existing work highlights the importance of both large-scale datasets and multimodal modeling for UI understanding. However, there remains a clear gap in datasets that combine real-world diversity with high-quality, expert-driven annotations at the element level. This motivates the need for more reliable and semantically accurate datasets to support the next generation of UI understanding systems.

\section{Methodology}

\begin{figure*}[t]
\centering
\includegraphics[width=0.95\textwidth]{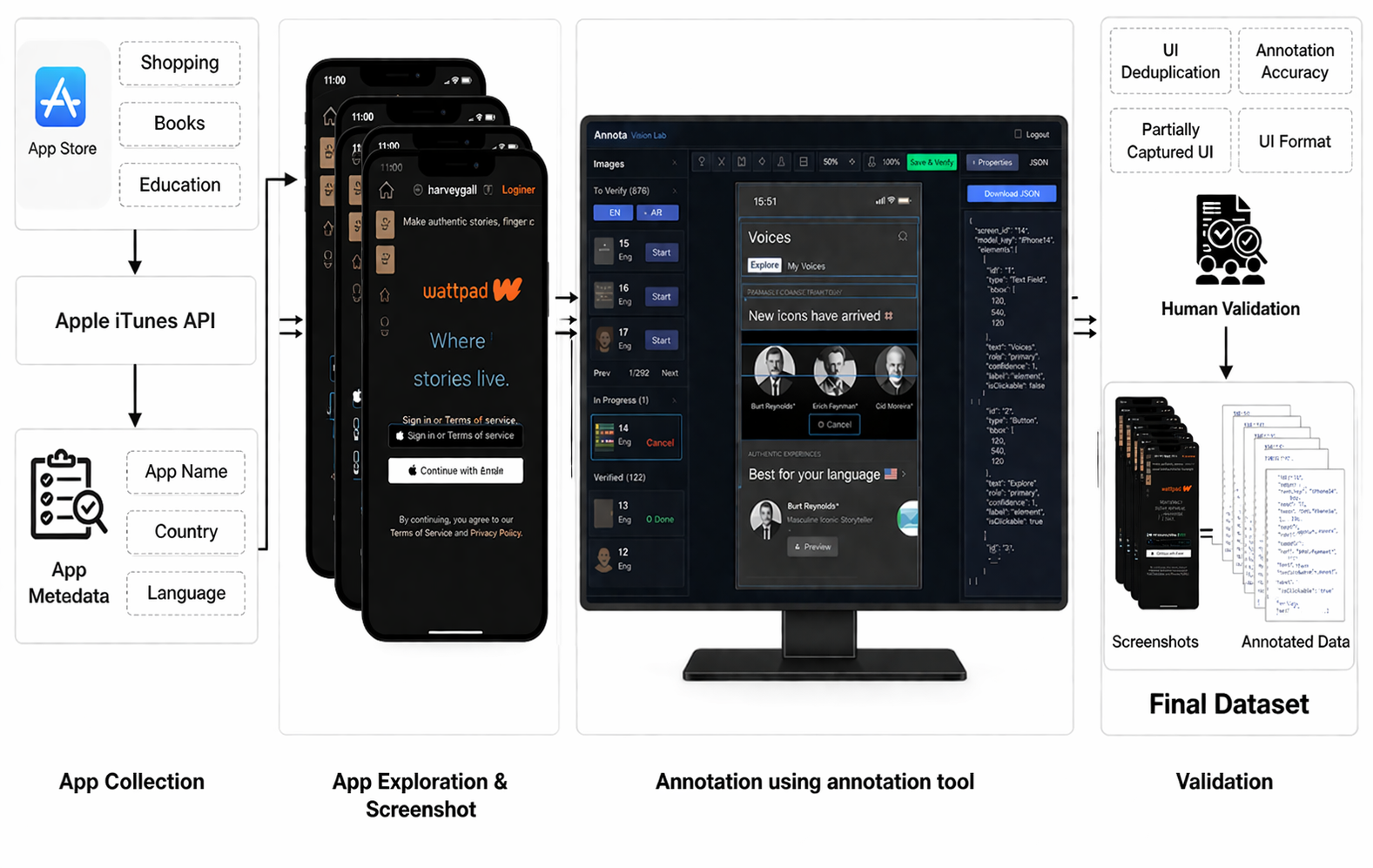}
\caption{Overview of the dataset construction workflow. Real-world iOS apps are selected using App Store metadata and the iTunes API, then manually explored to capture representative screenshots. The screenshots are annotated with bounding boxes and UI element labels using the custom annotation tool. Finally, the annotations are validated for accuracy, consistency, and completeness to produce the final dataset.}
\label{fig:dataset_construction}
\end{figure*}

This section presents the overall methodology as shown in Figure~\ref{fig:dataset_construction} adopted in this work which is describe in below figure including dataset statistics, annotation pipeline, annotation tool design, task formulation, and evaluation protocol. The goal is to construct a high-quality dataset and a reliable benchmark for mobile UI understanding.s

\subsection{Dataset}

\textbf{MUIAnno} consists of a diverse collection of mobile user interface (UI) screens obtained from real-world applications. A total of 38 applications were selected from the Apple App Store across multiple categories, including e-commerce, education, finance, social media, and other domains. This selection was made to capture a broad range of interface styles, layout structures, and interaction patterns commonly found in modern mobile applications.

The applications are drawn from 17 official App Store categories, which are grouped into 8 higher-level categories to provide a clearer and more consistent representation of application domains. From these applications, we collected 1,000 UI screens through manual exploration. A summary of the dataset statistics is provided in Table~\ref{tab:dataset_summary}. Each application contributes approximately 20–30 screens depending on its complexity and functionality. The collected screens represent different stages of user interaction, including onboarding flows, home interfaces, navigation menus, content browsing, and input-driven interfaces such as search and form entry. This diversity ensures that the dataset reflects realistic usage scenarios rather than isolated or synthetic interface samples. The dataset contains more than 27,367 annotated UI element instances, indicating dense annotation coverage across all screens. It is important to note that these annotations correspond to repeated occurrences of UI components across screens rather than unique element types. The annotation process follows a predefined taxonomy consisting of 36 UI element classes, derived from a structured annotation guideline to ensure consistency and semantic clarity. These classes span a wide range of UI components, including interactive elements (e.g., buttons, text fields, switches), visual elements (e.g., images, icons, illustrations), and structural or navigational elements (e.g., tab bars, dialogs, containers).

The selection of these 36 UI element classes was carried out through a systematic process combining empirical analysis and expert validation. Initially, a broad set of candidate UI components was identified by examining recurring design patterns across the collected applications. To further support this process, we referred to publicly available UI/UX design repositories such as Mobbin, which curate large collections of real-world iOS application interfaces for design exploration and research \citep{noauthor_discover_nodate}. This allowed us to identify commonly used interface elements and ensure that the taxonomy reflects contemporary mobile design practices. The initial set of candidate elements was subsequently refined through iterative expert review and annotation trials. During this stage, each element category was evaluated based on three key criteria: (1) \textit{coverage}, ensuring that frequently occurring UI components across diverse applications are included; (2) \textit{distinctiveness}, ensuring that each class represents a visually and semantically unique element to reduce ambiguity; and (3) \textit{annotation reliability}, ensuring that annotators can consistently identify and label elements based on clear visual cues. Elements that were ambiguous, infrequent, or difficult to annotate consistently were excluded or merged, while commonly observed and functionally meaningful components were retained. Furthermore, the taxonomy is aligned with established UI/UX design patterns and formalized through a structured annotation guideline that standardizes element definitions, labeling rules, and handling of complex cases such as nested components. This design ensures that the dataset is both practically annotatable and suitable for downstream tasks such as UI element detection and multimodal reasoning.

\begin{table}[h]
\centering
\caption{Summary of dataset statistics}
\label{tab:dataset_summary}
\begin{tabular}{lc}
\toprule
\textbf{Property} & \textbf{Value} \\
\midrule
Number of applications & 38 \\
Total UI screens & 1,000 \\
Application categories & 8 \\
UI element classes & 36 \\
Total annotated elements & 27,367 \\
Annotators & 6 \\
Annotation resolution & 828 $\times$ 1792 \\
Correction rate & 14\% \\
\bottomrule
\end{tabular}
\end{table}

The annotations capture fine-grained UI structure by labeling each visible element with tight bounding boxes. Nested elements (e.g., icons within buttons) are annotated separately to preserve their relationships, resulting in dense and structured annotations. The dataset is relatively balanced across application categories, with minor variations due to differences in app complexity. This diversity supports more robust and generalizable UI understanding.

The complete list of UI element classes is provided in Table~\ref{tab:ui_elements}.

\subsection{Annotation Process}

\begin{figure*}[t]
\centering
\includegraphics[width=0.85\textwidth]{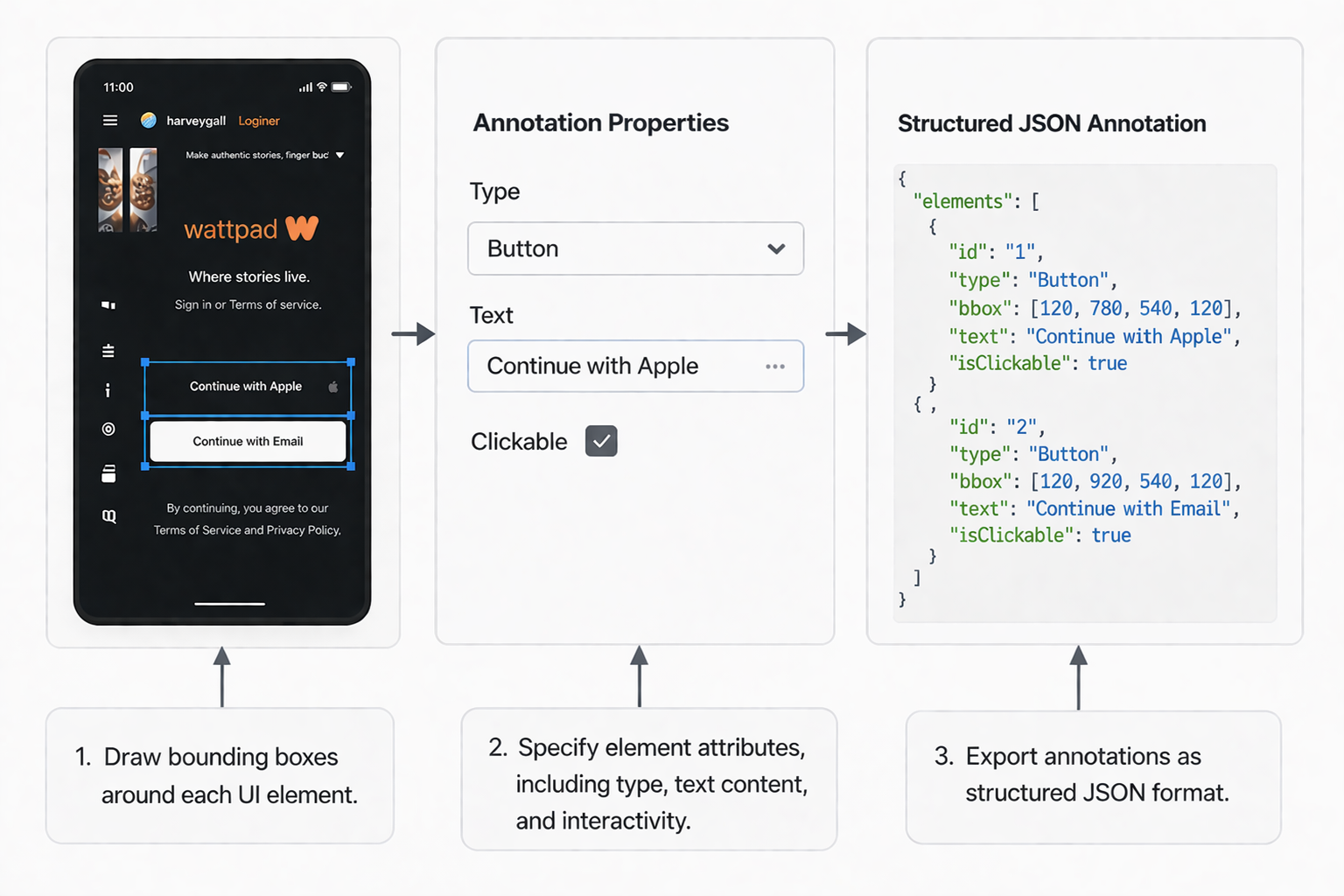}
\caption{Overview of the annotation pipeline. Annotators draw bounding boxes around UI elements, specify element attributes such as type, text content, and interactivity, and export the final annotations in structured JSON format.}
\label{fig:annotation_pipeline}
\end{figure*}

The annotation pipeline is designed to ensure consistent and semantically accurate labeling of UI elements across the dataset. An overview of the workflow is shown in Figure~\ref{fig:annotation_pipeline}. The overall design of the pipeline is informed by prior work on mobile UI dataset construction and screenshot-based UI understanding, where the quality of collected screens and the reliability of annotations are critical to downstream modeling performance \citep{deka_rico_2017, baechler_screenai_2024}.

The process begins with the collection of applications from the Apple App Store using the official iTunes platform. Specifically, the list of applications is first obtained through the iTunes Search API \citep{apple_itunes_search_api}, after which the selected applications are manually downloaded from the App Store. Each application is then manually explored to identify representative UI states, with the goal of capturing diverse stages of user interaction such as onboarding, navigation, content browsing, and input-driven screens. This human-driven exploration strategy is intended to preserve realistic interface states and interaction diversity, in contrast to purely automated collection pipelines \citep{gao_mobileviews_2024, feng_mud_2024}. Screenshots are collected using an iPhone 11 device to ensure consistency in visual appearance, resolution, and rendering behavior across all applications. All screens are captured at a fixed resolution of \textbf{828 $\times$ 1792 pixels}, ensuring a consistent visual format while retaining sufficient detail for fine-grained annotation.

All collected screens are uploaded to a custom web-based annotation tool, where annotators manually draw bounding boxes around visible UI elements and assign labels based on a predefined taxonomy. Each element is annotated individually, with particular attention to tightly enclosing the visual boundaries of the component. In assigning labels, annotators consider both the visual appearance of the element and its functional role within the interface. This is important because mobile UI understanding tasks often depend on both perceptual and semantic cues, as also reflected in prior work on widget-level annotation and screen-level semantic understanding \citep{wang_screen2words_2021, baechler_screenai_2024}.

To maintain consistency across annotators, a detailed annotation guideline is followed throughout the process. The guideline defines class descriptions, labeling rules, and procedures for handling ambiguous or structurally complex cases. Special attention is given to nested and composite UI elements. For example, when an icon appears within a button or text field, both the icon and the parent component are annotated separately. 

This strategy preserves structural relationships between elements and supports more fine-grained analysis of interface composition. The annotation process is carried out by annotators with prior UI/UX knowledge, as shown in Table~\ref{tab:annotators}, allowing decisions to be guided not only by visual boundaries but also by interface semantics. After the initial annotation stage, each screen undergoes a validation phase in which annotations are reviewed and refined. This includes correcting inaccurate labels, adjusting bounding boxes, and adding elements that may have been missed during the first pass. Such multi-stage refinement is especially important in dense mobile interfaces, where small components and closely packed layouts increase annotation difficulty.

\begin{table}[H]
\centering
\caption{UI element taxonomy used for annotation}
\label{tab:ui_elements}
\small
\setlength{\tabcolsep}{4pt}
\renewcommand{\arraystretch}{1.15}
\begin{tabularx}{\textwidth}{p{0.24\textwidth} p{0.23\textwidth} X}
\toprule
\textbf{Category} & \textbf{Element} & \textbf{Description} \\
\midrule

\multirow{10}{=}{Basic Elements}
& Accordion & Expandable or collapsible section revealing content \\
& Avatar & User profile image \\
& Badge & Small indicator showing count or status \\
& Banner & Wide message or notification bar \\
& Button & Clickable element triggering an action \\
& Card & Container grouping related content \\
& Checkbox & Multi-selection control \\
& Clickable Text & Text acting as a link or trigger \\
& Color Picker & UI for selecting colors \\
& Dropdown Menu & Expandable list of options \\
\midrule

\multirow{10}{=}{Visual and Informational}
& Illustration & Decorative or explanatory graphic \\
& Image & Photo or visual content \\
& Label & Text describing another element \\
& Loading Indicator & Spinner or progress indicator \\
& Logo & Brand or application symbol \\
& Map View & Embedded map component \\
& Plain Text & Static non-interactive text \\
& Status Dot & Indicator showing state \\
& Skeleton & Placeholder during loading \\
& Icon & Graphic symbol representing an action \\
\midrule

\multirow{8}{=}{Navigation and Layout}
& Side Navigation & Vertical navigation panel \\
& Switch & Toggle control \\
& Tab & Selectable section header \\
& Tab Bar & Horizontal navigation bar \\
& Table & Structured rows and columns \\
& Toolbar & Row of actions or tools \\
& Top Navigation Bar & Header containing title or actions \\
& Dialog & Modal interaction window \\
\midrule

\multirow{8}{=}{Interactive and Input}
& Radio Button & Single-choice selection control \\
& Search Bar & Input field for search queries \\
& Segmented Control & Group of selectable buttons \\
& Text Field & Input field for user text \\
& Date Picker & Input for selecting date \\
& Time Picker & Input for selecting time \\
& Floating Action Button & Prominent circular action button \\
& Gallery & Collection of images or media \\
\bottomrule
\end{tabularx}
\end{table}


\begin{table}[h]
\centering
\caption{Annotator expertise and experience}
\label{tab:annotators}
\begin{tabular}{llc}
\toprule
\textbf{Annotator} & \textbf{Expertise} & \textbf{Years of Experience} \\
\midrule
1 & UI/UX, visual design, UX research & 10 \\
2 & UI/UX, web design & 5 \\
3 & UI/UX & 3 \\
4 & UI/UX, graphic design & 2 \\
5 & UI/UX, graphic design & 2 \\
6 & UI/UX & 1 \\
\bottomrule
\end{tabular}
\end{table}

A final verification stage is then conducted to ensure consistency and completeness across the dataset. During this stage, annotations are checked for ambiguity, redundancy, and adherence to the predefined taxonomy. Any remaining inconsistencies are resolved through iterative review so that labeling remains uniform across screens. This multi-stage process results in high-quality annotations that are visually grounded, semantically consistent, and suitable for detailed UI understanding tasks.

\subsection{Annotation Tool}
\begin{figure*}[t]
\centering
\includegraphics[width=0.90\textwidth]{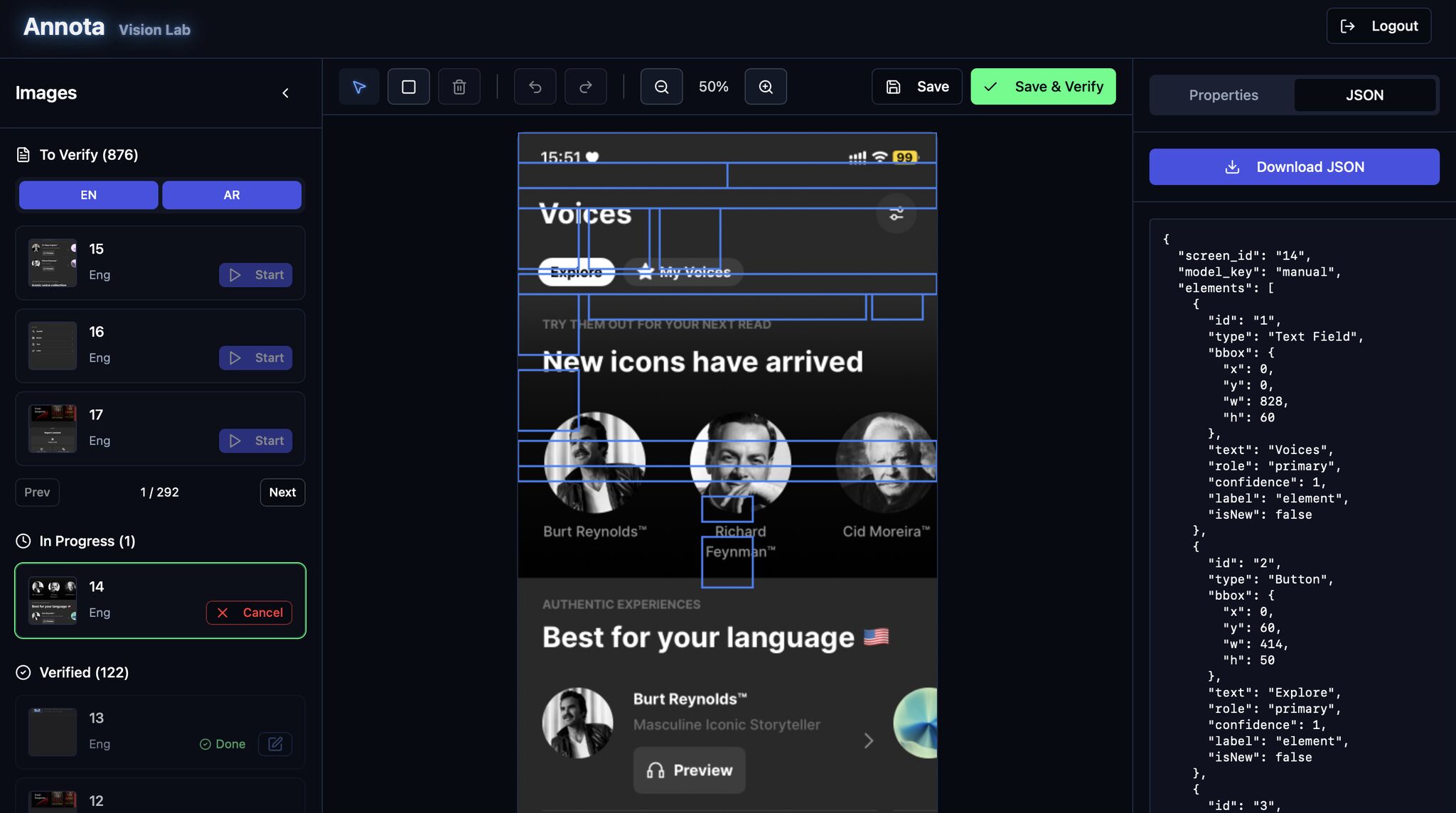}
\caption{Interface of the custom annotation tool used for labeling UI elements. Annotators can upload UI screens, draw bounding boxes, assign semantic labels, edit element attributes, and export the annotations in structured JSON format.}
\label{fig:annotation_tool}
\end{figure*}

To support efficient and consistent annotation, we developed a custom web-based annotation tool specifically designed for mobile UI labeling. The tool provides an interactive interface that allows annotators to upload UI screens, draw bounding boxes using a drag-and-drop mechanism, and assign semantic labels from a predefined set of UI element classes. Figure~\ref{fig:annotation_tool} shows the interface of the tool used in this study. The tool is designed to streamline the annotation process while maintaining consistency across annotators. It supports real-time editing, enabling annotators to refine bounding boxes and update labels as needed during the annotation process. This reduces annotation errors and allows for efficient handling of complex UI layouts. A key aspect of the tool is its support for structured data generation. All annotations are automatically exported in JSON format, where each UI element is represented by its bounding box coordinates and associated label. This structured representation ensures compatibility with downstream tasks such as UI element detection, layout analysis, and multimodal reasoning.

The \href{https://annota-nine.vercel.app}{annotation tool} is publicly accessible online, allowing reproducibility of the annotation process and enabling further extension of the dataset by the research community.

\subsection{Task Formulation}

To evaluate the proposed dataset, we adopt a prompt-based benchmarking framework using multimodal large language models (LLMs) accessed through publicly available APIs. Rather than training task-specific models, our goal is to assess how effectively general-purpose systems can perform fine-grained UI understanding under a unified evaluation setting. We formulate the task as \textit{UI element extraction}. Given a UI screenshot and a structured instruction prompt, the model is required to generate a structured JSON output describing all visible UI elements. Each element is represented by its semantic category and corresponding bounding box coordinates. This formulation is directly aligned with the dataset annotation schema, enabling consistent and interpretable comparison between model predictions and ground truth annotations. Unlike conventional object detection pipelines, no additional training or fine-tuning is performed. All models operate in a zero-shot or few-shot setting, relying solely on their pretrained multimodal capabilities. This design reflects practical deployment scenarios, where models are expected to interpret previously unseen interfaces without supervision, and allows the benchmark to focus on intrinsic reasoning and grounding ability.

To ensure fair comparison across different systems, we adopt a standardized prompt-based interface. The input instructions and expected output format are kept consistent for all models, minimizing variability introduced by prompt design. In addition, model outputs are constrained to a predefined JSON schema that matches the dataset annotation format, ensuring structural consistency and enabling reliable evaluation across different systems. The prompt template is fixed across all experiments and provided in the appendix to support reproducibility. Since the evaluation relies on API-based models, whose internal configurations may evolve over time, we enforce deterministic generation settings and consistent prompting strategies. This helps maintain stability and comparability of results across experiments.

\FloatBarrier

\subsection{Evaluation Protocol}\label{sec:evaluation}

Model performance is evaluated using a combination of spatial and semantic metrics to jointly assess localization accuracy and classification correctness. Since the task requires structured prediction of UI elements, the evaluation protocol must capture both geometric alignment and semantic consistency. Spatial alignment is measured using Intersection over Union (IoU), which measures the overlap between a predicted bounding box and its corresponding ground-truth bounding box. IoU is defined as:

\begin{equation}
IoU = \frac{B_p \cap B_g}{B_p \cup B_g}
\end{equation}

where $B_p$ denotes the predicted bounding box and $B_g$ denotes the ground-truth bounding box. A predicted UI element is considered correctly localized if its IoU with a ground-truth element is at least 0.5. For matching, each predicted bounding box is assigned to the ground-truth element with the highest overlap under a one-to-one matching constraint. This prevents multiple predictions from being matched to the same ground-truth element. Detection performance is evaluated using precision, recall, and F1-score. Precision measures how many predicted elements are correct, while recall measures how many ground-truth elements are successfully detected. These metrics are computed as:

\begin{equation}
Precision = \frac{TP}{TP + FP}
\end{equation}

\begin{equation}
Recall = \frac{TP}{TP + FN}
\end{equation}

where $TP$ represents correctly detected UI elements, $FP$ represents incorrectly predicted elements, and $FN$ represents missed ground-truth elements. A prediction is counted as a true positive only when both conditions are satisfied: the bounding box meets the IoU threshold and the predicted label matches the ground-truth category. To provide a balanced measure between precision and recall, we also report the F1-score, which is widely used in detection and classification tasks when both false positives and false negatives are important \citep{powers2011evaluation}. In our setting, this is particularly relevant because models may either over-predict UI elements or miss small and densely arranged components. The F1-score is computed as:

\begin{equation}
F1 = 2 \times \frac{Precision \times Recall}{Precision + Recall}
\end{equation}

In addition to localization accuracy, we evaluate semantic correctness by checking whether the predicted UI element type matches the ground-truth label. Since the models generate structured JSON outputs, malformed outputs, incorrect labels, missing elements, and duplicate predictions directly affect the final evaluation. Overall, this protocol reflects realistic UI understanding scenarios, where accurate bounding-box localization and correct semantic interpretation are both necessary.

\section{Benchmark Experiments}

In addition to presenting the dataset, we evaluate its effectiveness through a benchmark designed for fine-grained UI understanding. The benchmark assesses how well modern multimodal large language models (LLMs) can interpret mobile interfaces and generate structured representations of UI elements. All experiments are conducted on the complete dataset consisting of 1,000 UI screens to ensure comprehensive and reliable evaluation. Rather than training task-specific models, this setup evaluates the ability of general-purpose systems to perform UI element extraction directly from screenshots using prompt-based reasoning.

\subsection{Task Definition}

We define a task referred to as \textit{UI Element Extraction}, where the objective is to identify and describe all visible UI components within a given mobile screen. Each model is provided with a UI image together with a structured instruction prompt that specifies the expected output format, along with a representative example to guide generation. The expected output is a structured JSON representation that includes the semantic type of each UI element and its corresponding bounding box coordinates. This formulation is consistent with the annotation schema introduced in the dataset, enabling direct and interpretable comparison between model predictions and ground truth annotations. The task is designed to reflect practical deployment scenarios, where models are required to interpret complete UI layouts without task-specific training. It provides a unified setting for evaluating both spatial localization and semantic understanding of interface elements.

\subsection{LLM-based Annotation Pipeline}

To support this task, we adopt a prompt-driven annotation pipeline implemented through an automated workflow. For each UI screen, the image is provided to the model along with a fixed instruction prompt and a reference example. The model is required to generate a complete set of UI elements in a structured JSON format aligned with the dataset annotation schema. The workflow is implemented using the \textit{n8n} automation framework~\citep{n8n}, deployed locally in a Docker-based environment. This setup enables the orchestration of API-based model inference, input preparation, and output collection within a unified and reproducible pipeline. Each UI screen is processed sequentially, where requests are sent to the selected models and responses are collected in a standardized format.

To ensure consistency and reduce output variability, model responses are constrained using a predefined JSON schema that enforces the structure of the generated annotations. This constraint improves the reliability of outputs and simplifies downstream evaluation by ensuring compatibility with the ground truth annotation format. All models are evaluated under identical conditions using a fixed prompt template and reference example. The generation process is configured with deterministic settings (e.g., low temperature), minimizing stochastic variation and enabling stable comparison across models. This design ensures that observed performance differences primarily reflect model capabilities rather than variations in prompting or sampling.

\subsection{Evaluated Models}

We evaluate five representative multimodal large language models (LLMs) for the UI element extraction task. The evaluated models include three closed-source systems, namely OpenAI GPT-5.4, Anthropic Claude Opus 4.6, and Google Gemini 3.1 Pro (Preview), as well as two open-source multimodal models, Gemma-4-31B-IT and meta-llama/Llama-4-Scout. All models are accessed through API-based inference to keep the experimental setup consistent across both closed-source and open-source systems.

All models are selected based on their strong multimodal capabilities and their ability to perform vision-language tasks. In particular, these models can process UI screenshots and generate structured outputs, which makes them suitable for prompt-based UI element extraction. The closed-source models are included because they are among the leading proprietary multimodal systems, while the open-source models are included to provide accessible and reproducible baselines. In addition, the selected models are commonly discussed or listed among high-performing systems in public evaluations and leaderboards, such as the LLM Arena benchmark~\citep{noauthor_llm_arena_2024}, which reports model performance based on human preference and overall capability. The open-source models are included to provide additional reproducible baselines. Although they are evaluated through APIs in this study for consistency with the other models, they can also be downloaded and deployed locally by researchers. This makes them useful for future reproduction, inspection, and extension of the benchmark. Their inclusion helps assess how accessible open-source multimodal models perform on fine-grained UI understanding compared with leading proprietary systems. Together, the selected models cover different model families, providers, and availability settings. This makes the benchmark more balanced, since it includes both closed-source systems and open-source alternatives while keeping the evaluation protocol consistent. For all models, the same UI screenshot and instruction prompt are used as input, and the outputs are constrained to a unified JSON format. This ensures that the comparison focuses on each model's ability to accurately localize UI elements and assign correct semantic labels under consistent evaluation conditions.

\subsection{Implementation Details}

All experiments are conducted using the unified workflow described above. To maintain consistency, both closed-source and open-source models are evaluated through API-based access. Although the open-source models can be downloaded and deployed locally, API-based inference is used in this study to ensure that all models are tested under the same prompt-based setting, input format, and output schema. For every model, we use the same prompt structure, reference example, and JSON output schema. The experiments are performed on the full dataset of 1,000 UI screens, covering diverse application categories, layouts, and interface styles. Since the experiments are conducted in a zero-shot setting without any task-specific training or fine-tuning, no dataset split is required. All screens are therefore used directly for evaluation. Model predictions are assessed using the IoU-based matching protocol, label correctness, precision, recall, and F1-score metrics defined in Section ~\ref{sec:evaluation}.

\subsection{Reproducibility Considerations}

Since all evaluated models are accessed through API-based inference in this study, their behavior may change over time due to updates in the underlying systems. To reduce this effect, we use fixed prompts, schema-constrained outputs, and deterministic generation settings across all experiments. In addition, all evaluations are conducted on the full dataset, which helps improve the robustness of the reported results and reduces potential sampling bias.

Exact reproducibility is still challenging for API-based evaluations because model versions, serving configurations, and backend updates may not always remain fixed. This limitation is particularly relevant for closed-source systems. To partially address this issue, we include open-source multimodal models in the benchmark. Although they are evaluated through APIs in this study for consistency, these models can also be downloaded, deployed, inspected, and re-evaluated by other researchers under similar settings. Overall, the combination of fixed evaluation conditions, full-dataset testing, and the inclusion of open-source models strengthens the reproducibility of the benchmark while still allowing comparison with leading proprietary systems.

\section{Results and Discussion}

\begin{figure*}[t]
\centering
\includegraphics[width=0.85\textwidth]{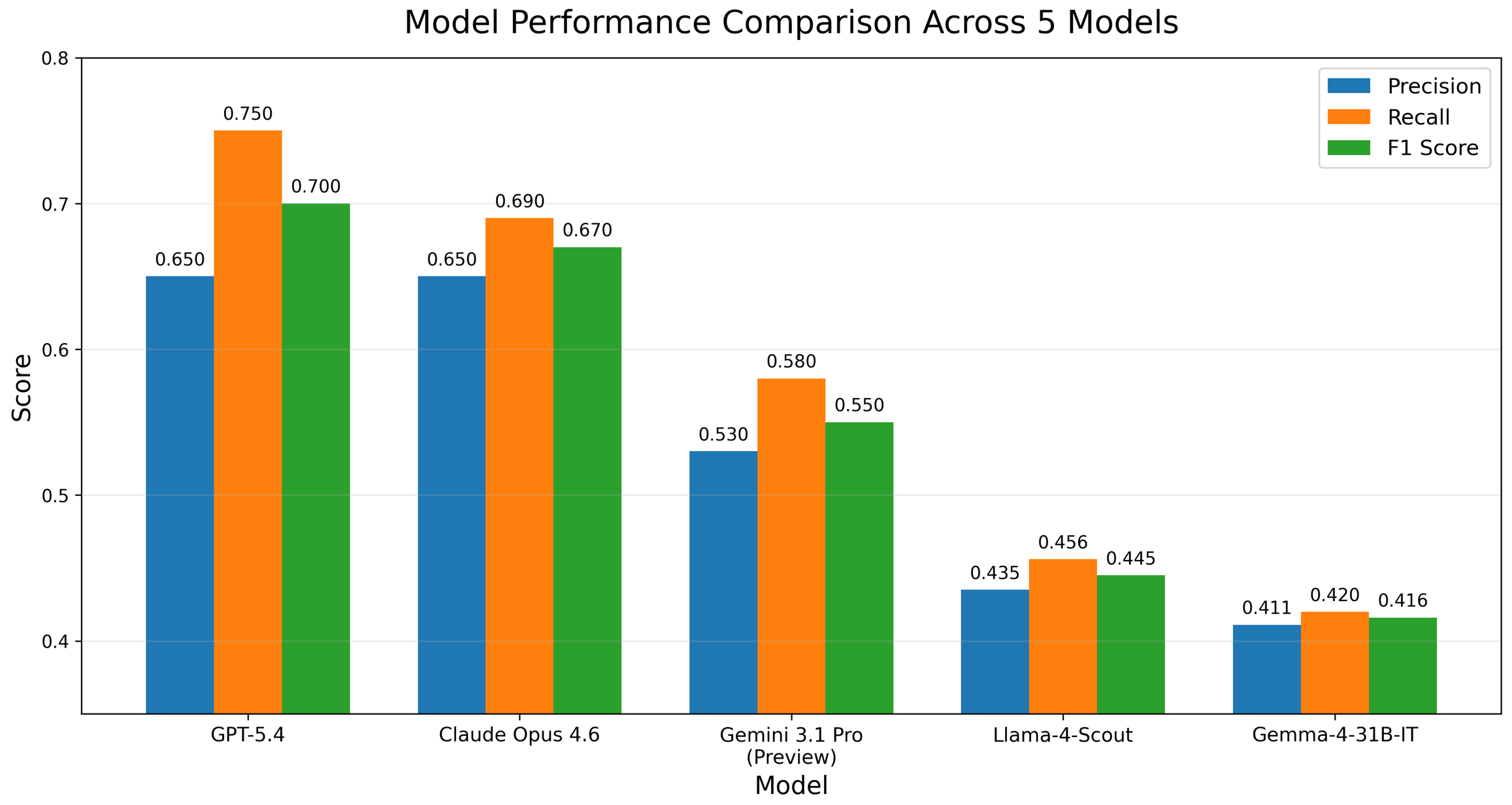}
\caption{Comparison of precision, recall, and F1-score across evaluated multimodal LLMs.}
\label{fig:model_comparison}
\end{figure*}

This section presents the experimental results of the evaluated multimodal LLMs on the UI element extraction task and discusses their performance in terms of localization accuracy, semantic correctness, and overall detection quality. The analysis focuses on comparing closed-source and open-source models under the same prompt-based evaluation setting. 
Table~\ref{tab:llm_results} summarizes the precision, recall, and F1-score of each model, while Figure~\ref{fig:model_comparison} provides a visual comparison of their performance.

\begin{table}[h]
\centering
\caption{Performance comparison of multimodal LLMs on UI element extraction}
\label{tab:llm_results}
\begin{tabular}{lccc}
\toprule
\textbf{Model} & \textbf{Precision} & \textbf{Recall} & \textbf{F1-score} \\
\midrule
GPT-5.4 & 0.65 & 0.75 & 0.70 \\
Claude Opus 4.6 & 0.65 & 0.69 & 0.67 \\
Gemini 3.1 Pro (Preview) & 0.53 & 0.58 & 0.55 \\
meta-llama/Llama-4-Scout & 0.435 & 0.456 & 0.445 \\
Gemma-4-31B-IT & 0.411 & 0.420 & 0.416 \\
\bottomrule
\end{tabular}
\end{table}

The results show that current multimodal LLMs can perform UI element extraction from screenshots in a prompt-based setting without task-specific training. However, the performance varies noticeably across models. Among all evaluated models, GPT-5.4 achieves the best overall result, with the highest F1-score of 0.70. This is mainly supported by its strong recall of 0.75, indicating that it identifies a larger proportion of UI elements present in the screenshots. Claude Opus 4.6 achieves a similar precision score of 0.65 but slightly lower recall of 0.69, resulting in an F1-score of 0.67. This suggests that Claude produces predictions with a comparable level of correctness, but it misses more UI elements than GPT-5.4. Its relatively balanced precision and recall indicate stable performance, although with slightly lower coverage. Gemini 3.1 Pro (Preview) obtains an F1-score of 0.55, with precision of 0.53 and recall of 0.58. Compared with GPT-5.4 and Claude Opus 4.6, its lower scores suggest weaker performance in both localization accuracy and element coverage. This indicates that Gemini struggles more with fine-grained UI element extraction, especially when screens contain many small or closely arranged components.

The two open-source models show lower performance than the proprietary models. meta-llama/Llama-4-Scout achieves a precision of 0.435, recall of 0.456, and F1-score of 0.445, while Gemma-4-31B-IT obtains a precision of 0.411, recall of 0.420, and F1-score of 0.416. Although these results are lower, the open-source models provide useful baselines because they can be downloaded, deployed, and re-evaluated by other researchers. Their inclusion also helps show the current performance gap between leading proprietary multimodal systems and accessible open-source alternatives on fine-grained UI understanding. Across the evaluated models, recall is generally higher than or close to precision, suggesting that models often attempt to detect a broad set of UI elements but may also generate incorrect or imprecise predictions. GPT-5.4 shows the strongest recall, which indicates better coverage of visible UI elements. In contrast, the open-source models achieve lower recall, suggesting that they miss more elements, particularly smaller components or elements with less visual salience. Performance also depends on the visual characteristics of UI elements. Larger and more visually clear components, such as buttons, text blocks, cards, and images, are generally easier for the models to detect. Smaller elements, such as icons, status indicators, and densely packed interface components, are more frequently missed or inaccurately localized. This shows that current multimodal models are better at capturing coarse UI layout structure than precise element-level details. Despite the promising results, all models still face limitations in spatial grounding. In many cases, predicted bounding boxes only partially overlap with the corresponding ground truth elements, or extra elements are generated that do not correspond to valid annotations. These errors become more common in complex screens with dense layouts, nested components, or visually similar UI elements.

The use of schema-constrained JSON outputs improves the structural consistency of model predictions and makes the outputs easier to evaluate against the dataset annotations. However, this constraint does not fully solve errors related to incorrect labels, missing elements, or inaccurate bounding box placement. Therefore, both visual grounding and semantic classification remain important challenges for UI element extraction. From a dataset perspective, these results highlight the value of high-quality, fine-grained annotations. The dense annotation structure makes it possible to evaluate not only whether a model understands the general screen layout, but also whether it can identify individual UI components accurately. This is especially important for benchmarking multimodal models on realistic mobile interfaces, where small elements and nested structures are common. Overall, the findings indicate that multimodal LLMs can perform UI element extraction directly from screenshots, but their performance remains limited by layout complexity, element scale, and spatial precision. The stronger performance of proprietary models shows the capability of current leading systems, while the open-source baselines provide a useful reference point for reproducible future research. The proposed benchmark therefore offers a practical framework for analyzing the strengths and weaknesses of multimodal models in fine-grained mobile UI understanding.

\section{Conclusion}

This work introduced \textbf{MUIAnno}, a new expert-annotated dataset for mobile UI understanding, consisting of real-world iOS screens labeled at the element level by UI/UX experts. MUIAnno provides structured annotations that capture both the visual appearance and functional roles of UI components, making it suitable for UI element extraction, layout analysis, and multimodal interface understanding. We also presented a prompt-based benchmark using proprietary and open-source multimodal LLMs under a consistent API-based evaluation setting. The results show that current models can generate structured UI element predictions from screenshots without task-specific training, but they still face challenges in detecting small or densely arranged components, producing accurate bounding boxes, and assigning consistent semantic labels in complex layouts. Overall, MUIAnno and the associated benchmark provide a useful foundation for evaluating fine-grained mobile UI understanding systems. Future work may extend the dataset to more dynamic interaction patterns and explore stronger integration between multimodal models and structured UI representations to improve spatial grounding and robustness.

\section*{Disclosure Statement}

The author reports no potential conflict of interest.

\section*{Funding}

No funding was received for this work.

\section*{Data Availability Statement}

The MUIAnno dataset is publicly available at: \url{https://huggingface.co/datasets/atharparvezce/MUIAnno-Mobile-UI-Dataset}. The annotation tool is publicly available at: \url{https://annota-nine.vercel.app}.

\bibliographystyle{apalike}
\bibliography{references}

\end{document}